\documentclass{article} 
\usepackage{nips13submit_e,times}

\usepackage{cite}
\usepackage[hyphens]{url}

\usepackage{graphicx}
\usepackage[space]{grffile}
\usepackage{latexsym}
\usepackage{amsfonts,amsmath,amssymb}
\usepackage{url}
\usepackage[utf8]{inputenc}
\usepackage{fancyref}
\usepackage{hyperref}
\hypersetup{colorlinks=false,pdfborder={0 0 0},}
\usepackage{textcomp}
\usepackage{longtable}
\usepackage{multirow,booktabs}

\title{Progmosis:\\Evaluating Risky Individual Behavior During Epidemics Using
Mobile Network Data}

\author{Antonio Lima\\ Massachussets Institute of Technology (MIT)\\University of Birmingham, UK\\  \And Veljko Pejovic\\ University of Birmingham, UK\\ \And Luca Rossi\\ University of Birmingham, UK\\ \And Mirco Musolesi\\ University of Birmingham, UK\\ \And Marta Gonzalez\\ Massachusetts Institute of Technology (MIT)\\}

\nipsfinalcopy 

\begin{document}

\bibliographystyle{unsrt}

\newcommand{\fix}{\marginpar{FIX}}
\newcommand{\new}{\marginpar{NEW}}

\maketitle

\begin{abstract}
The possibility to analyze, quantify and forecast epidemic outbreaks is
fundamental when devising effective disease containment strategies.
Policy makers are faced with the intricate task of drafting
realistically implementable policies that strike a balance between risk
management and cost. Two major techniques policy makers have at their
disposal are: epidemic modeling and contact tracing. Models are used to
forecast the evolution of the epidemic both globally and regionally,
while contact tracing is used to reconstruct the chain of people who
have been potentially infected, so that they can be tested, isolated and
treated immediately. However, both techniques might provide limited
information, especially during an already advanced crisis when the need
for action is urgent.

In this paper we propose an alternative approach that goes beyond
epidemic modeling and contact tracing, and leverages behavioral data
generated by mobile carrier networks to evaluate contagion risk on a
per-user basis. The individual risk represents the loss incurred by not
isolating or treating a specific person, both in terms of how likely it
is for this person to spread the disease as well as how many secondary
infections it will cause. To this aim, we develop a model, named
\emph{Progmosis}, which quantifies this risk based on movement and
regional aggregated statistics about infection rates. We develop and
release an open-source tool that calculates this risk based on cellular
network events. We simulate a realistic epidemic scenarios, based on an
Ebola virus outbreak; we find that gradually restricting the mobility of
a subset of individuals reduces the number of infected people after 30
days by 24\%.

While these results are promising, it is important to underline the fact
that this is only an initial foundational work and to stress some key
points. First, this paper focuses on a theoretical model, rather than on
its actual translation into a real-world system. In particular,
centralized deployments of this model would pose several ethical
questions, as they would require access to user data. Decentralized
deployments for which user mobility data never leaves the mobile device
of a user are possible and should be preferred, as they fully protect
user privacy. Second, results are generated from computer-based
simulations, under specific assumptions. Social factors and technical
difficulties might greatly affect results obtained in the real world.
Third, this risk-assessment tool is not designed specifically for
implementing containment measures based on mobility restrictions. For
example, it could be used to advise users about the most appropriate
behavior given his/her risk profile (e.g., willingly change own
behavior, see a doctor, and similar); users would finally choose whether
to follow the advice or not. Finally, the simulations were run on data
call records from a country that is according to WHO
Ebola-free~\cite{who_senegal_2014}, and this work has not been
commissioned neither by Orange nor by any other entity for preparation
to a real-world disease outbreak.

\end{abstract}

\section{Introduction}
The world is facing a number of severe healthcare challenges and, indeed, the recent Ebola outbreak seems one of the most worrisome and urgent. Mr David Nabarro, Special Envoy of the UN Secretary-General, said at an informal UN meeting that he had never encountered a challenge like Ebola in 35 years of his professional life: ``This outbreak has moved out of rural areas and it's coming to towns and cities. It's no longer just affecting a very well-defined location, it's affecting a whole region and it's now impacting the whole world"\footnote{\url{http://webtv.un.org/watch/david-nabarro-ebola-virus-outbreak-general-assembly-informal-meeting-69th-session-10-october-2014/3832613824001}}.

Nowadays transportation systems make it possible for people to travel easily across a country and across the globe, but, unfortunately, they make that possible for diseases too. The spread of diseases is facilitated by today's rich transportation networks that enable human disease carriers to quickly move across distant regions~\cite{Merler2010}. In this context, drastic measures like banning transportation to disease-affected areas are difficult to implement, have a high cost and are actually believed to worsen the outbreak~\cite{ebola_forbes}~\cite{Meloni2011}. The need for smaller, targeted interventions matches the increasing availability of large-scale data, especially coming from mobile networks. The benefit of mobile-phone records to combat quickly-spreading diseases like Ebola is unquestionable~\cite{economist_call_2014}.

When an outbreak becomes global, an infected person can be found anywhere, in cities as well as rural areas, and regardless of country boundaries; this might suggest that no place is really safe. However, we argue that some people and places are more exposed to the risk than others.

We propose to use such heterogeneity to our advantage and to use mobile networks to unveil such heterogeneity.
We envision a system that utilizes the data coming from mobile carriers and, where available, social networks and smartphones, to construct \textit{individual-based risk models}. The system can assess the risk associated with a person, primarily based on that person's mobility patterns and, optionally, on other demographic or behavioral indicators that can be inferred from the data. We would like to highlight the characterizing features of the proposed solution: first, it can use data that is readily available (such as cellphone carrier data), and second, it is be able to operate under uncertainty (it does not require the knowledge of the identity of the infected).

The risk model can be used in several real-world scenarios, especially when a urgent response is required. Thus, the model can be used to answer the following questions. Who should be tested early for signs of the disease, and possibly put into quarantine if positive, given that vaccinations can be produced and performed with a certain rate? Who should get vaccinated first? Who should receive information about prevention, for example by means of text messages? All these scenarios describe individual-based interventions that are very hard to administer quickly over large populations. This model can prioritize the people to be targeted with the intervention sooner rather than later.

\section{Motivation}

\textbf{People behavior is highly heterogeneous.} Existing epidemic models are based on analyses conducted at population level to assess how infectious a disease is, based on the basic reproductive ratio $r_0$, i.e., the average number of secondary cases generated by a single infected person. However, several studies have concluded that spreading processes are usually highly heterogeneous and that some individuals remain responsible for a large proportion of the spreading. The presence of these influential spreaders has been investigated for generic networks~\cite{kitsak_identification_2010}, as well as in epidemics processes. Superspreading seems to be a common feature of the spread of diseases and targeted individual-based control measures are much more effective than population-wide measures, as reported by Lloyd-Smith et al.~\cite{lloyd-smith_superspreading_2005}. For this reason, identifying superspreaders is extremely important in order to contain epidemics.

\textbf{Existing techniques, such as contact tracing, are not sufficient.} Moreover, efforts in fighting disease outbreaks mainly focus on contact tracing techniques, as it is happening for Ebola~\cite{murphy_contact_2014}. Contact tracing works by finding all the people who have been in contact with an infected person, and then interviewing, monitoring, isolating them when necessary.
The process is repeated for everyone who is found to be infected. While contact tracing can be effective, it has some drawbacks. First of all, information provided by people might be subject to errors, due to fear, shame, faulty memory or other reasons. Secondly, contact tracing needs time: contact tracing only starts when a person is already diagnosed with the disease, or at least shows symptoms. Tracing the contacts also takes time: if the disease has an asymptomatic phase or is highly infective, the contacts might be likely to have infected others before they are traced.

\textbf{Localization techniques have already been used successfully during critical scenarios.} Recently, Nigeria also resorted to GPS technology to improve, scale up and speed up contact tracing, repurposing GPS devices used for polio vaccinations~\cite{fasina2014transmission,gates_gis}. The huge effort of the country resulted in eradication of Ebola and Nigeria was declared "Ebola-free" by the WHO\footnote{\url{http://www.who.int/mediacentre/news/ebola/20-october-2014/en/}}. While this success story demonstrates how location tracking can be very useful during similar scenarios, the very same strategy could have not been used if the epidemic was in a more advanced state, i.e., if many more people had already been infected. For this reason, we believe it is very important to investigate the use of alternative systems that can provide coarser location tracking but for a large number of individuals.

\textbf{Medical treatment is scarce and costly.} For example, in the case of Ebola, although the disease is seen as a serious challenge by the whole world, vaccinations have to face serious technical and financial issues before being administered\footnote{\url{http://news.sciencemag.org/health/2014/10/leaked-documents-reveal-behind-scenes-ebola-vaccine-issues}}). When a commodity such as vaccinations is scarce, who should be given priority during vaccination?

\section{Risk Model}

In this section, we propose a method to quantify the risk associated with each person during an outbreak, depending on their mobility behavior, inferred from their phone-activity. Here we refer it as the risk model. Our goal is not the estimation of the individual cost (i.e., the chance of getting infected), but the cost that an entire community faces by not treating a specific person. Early testing, medical treatment, vaccination, quarantine of specific individuals might reduce cost sustained by the community at later time.

A general estimate of the total risk $R$ associated to a set of events $E$ is defined by:
\begin{equation}
R = \sum_E P_E \times L_E
\end{equation}
where $P_E$ and $L_E$ are the probability and the expected loss for each event, respectively~\cite{vapnik1998statistical}.

We bring this definition to the epidemiology domain by considering a scenario in which several geographic areas are assigned different values of time-varying contagion risk. The risk measures how likely it is for an individual to get infected in a region. As in common models of infectious diseases, we assume it is directly proportional to the fraction of infected people in the region and we also assume homogeneous mixing within the region. Similarly, we assume that the risk to infect a healthy individual is directly proportional to the fraction of susceptible people in the region.

By staying in a geographic area with a non-zero risk, a person will have some chances to get infected; the same person will also have a chance to infect someone else, increasing the risk of the geographic area. When moving between two or more areas, the person will affect the risk of these areas. We will not determine whether each person is in a susceptible, infective or recovered state. Instead, we will consider them in all the states and we will assess how risky their mobility behavior is.

In general, the way people transmit disease across geographic areas has been extensively studied in literature~\cite{balcan2009multiscale,Merler2010,bajardi2011human}. Most of the studies dealing with the effects of mobility on epidemic spreading usually make the assumption that the mobility patterns of individuals in a subpopulation are homogeneous~\cite{Colizza_2007}, while they are indeed highly heterogeneous~\cite{Merler2010,dalziel2013human}. This is particularly true for developing countries, where highly irregular and temporally unstructured contact patterns have been observed~\cite{vazquez2013using}.

We consider a disease that has contagion rate per contact $\beta$ (i.e., given a friendship between an infected and a susceptible person, a contagion will happen with rate $\beta$). Assuming the user $u$ spends $T_{u, l}$ fraction of his time in each location $l \in \mathcal{L}_u$ (hence, $\sum_i T_{u, i} = 1$) we define a time-dependant contagion risk:
\begin{equation}
C_u(t) = \beta \sum_{l, m \in \mathcal{L}} T_{u, l} T_{u, m} [ i_l(t) s_m(t) + i_m(t) s_l(t) ].
\end{equation}

where $i_l(t)$ and $s_l(t)$ refer to the of fraction of infected and susceptible population in location $l$ at time $t$, respectively. Note that now the probability of the event occurring, in this case, is the probability that a person becomes infected in a region, according to the time fraction spent there, while the expected loss is the number of people expected to be infected in another region, according to the time fraction spent there. As we do not know where the person might be infected, this formula accounts for all the combinations, which are assumed as equally likely. The maximum risk value, for a specific state of the network, is reached by an individual who equally spends his time in the region with the highest infected fraction of individuals and in the region with the highest susceptible fraction. We might calculate this normalized value but, for ranking purposes, it is not necessary, as it is a common factor; we can also ignore the rate $\beta$ for the same reason.

Our proposed model could be generalized by defining different risk classes depending on demographic indicators, which can be inferred from mobile data~\cite{Zhong_2013} or other behavioral indicators, such as those provided with the D4D-Dataset~\cite{de_montjoye_d4d-senegal:_2014}. It is important to emphasize that a real-world system that would use this model would require access to two types of data: global information about the outbreak, which is already available  (e.g., estimated number of infected people in various geographic regions); individual information about user mobility, which is sensitive information. For this reason, centralized deployments of the system might not be realizable, as user mobility might be unaccessible under local regulations and laws. On the other hand, decentralized deployments are to be preferred. In such deployments the mobility data never leaves the user device, the risk-profile is calculated on the phone and it is shown only to the user, who optionally chooses to follow tailored advices.

\section{Evaluation}
Next, we evaluate the effectiveness of the risk identification and containment model proposed above. We set up a realistic epidemic scenario and perform stochastic simulations, following an approach similar to that implemented in GLEaM~\cite{Balcan_2009}, while keeping track of the movement of individuals following the real traces found in the dataset. We use the SEIR model, where each individual can be in one of the following discrete states at any given time instant: susceptible (S), exposed (E), infected (I), permanently recovered or deceased (R). This model has been used for the 2002 seasonal influenza outbreak~\cite{Balcan_2009} and the 2014 Ebola outbreak~\cite{Althaus_2014}, among other outbreaks. It is described by the following set of equations:
\begin{eqnarray}
\frac{dS}{dt} &=& - \beta S(t) I(t)/N\\
\frac{dE}{dt} &=& \beta S(t) I(t) / N - k E(t) \\
\frac{dI}{dt} &=& k E(t) - \gamma I(t) \\
\frac{dR}{dt} &=& \gamma I(t)
\end{eqnarray}

We inform a spreading model with the realistic parameters taken from estimates of the 2014 Ebola outbreak in Sierra Leone~\cite{Althaus_2014}, as reported in Tab.~\ref{tab:params}.
Where $\sigma^{-1}$ and $\gamma^{-1}$ are the average durations of incubation and infectiousness, respectively. The transmission rate per day in absence of control interventions is $\beta$, and $r_{0}=\beta/\gamma$ is the basic reproduction number.
\begin{table}[h]
\centering
\begin{tabular}{|ll|}
$\sigma^{-1}$ & 5.3 [days] \\ 
$\gamma^{-1}$ & 5.61 [days] \\
$r_0$ & 2.53 \\
$\beta$ & 0.45
\end{tabular}
\label{tab:params}
\caption{Parameters assumed for the simulation.}
\end{table}

We simulate the epidemics in the following different contexts:

\begin{itemize}
\item in total absence of any treatment;
\item when treatment is given with rate $\xi$ per day and people given treatment are chosen randomly;
\item when treatment is given with rate $\xi$ per day to highest ranked people, according to the risk measure $C_u$.
\end{itemize}

For simplicity, in this paper we focus only on treatment that takes the form of travel restrictions, not allowing high-risk individuals to travel outside the metapopulation they are found when the treatment is applied. This is an extreme scenario, realistic only for diseases for which specific treatments or vaccinations are not available (e.g., Ebola virus). Without loss of generality, we can investigate the effects of vaccination and/or early treatment of people with higher-risk movement patterns. Since we use the same parameters for each metapopulation, and the treatment does not directly affect the epidemic process (i.e., it is not a vaccination or a cure) but only the movement of individuals, the local epidemic profiles will be similar and will be more or less shifted in time, depending on the travel fluxes. We will first show how much we can reduce synchronization by restricting the travel of high-risk individuals in a simple example.

\begin{figure}[h!]
\begin{center}
\includegraphics[width=0.7\columnwidth]{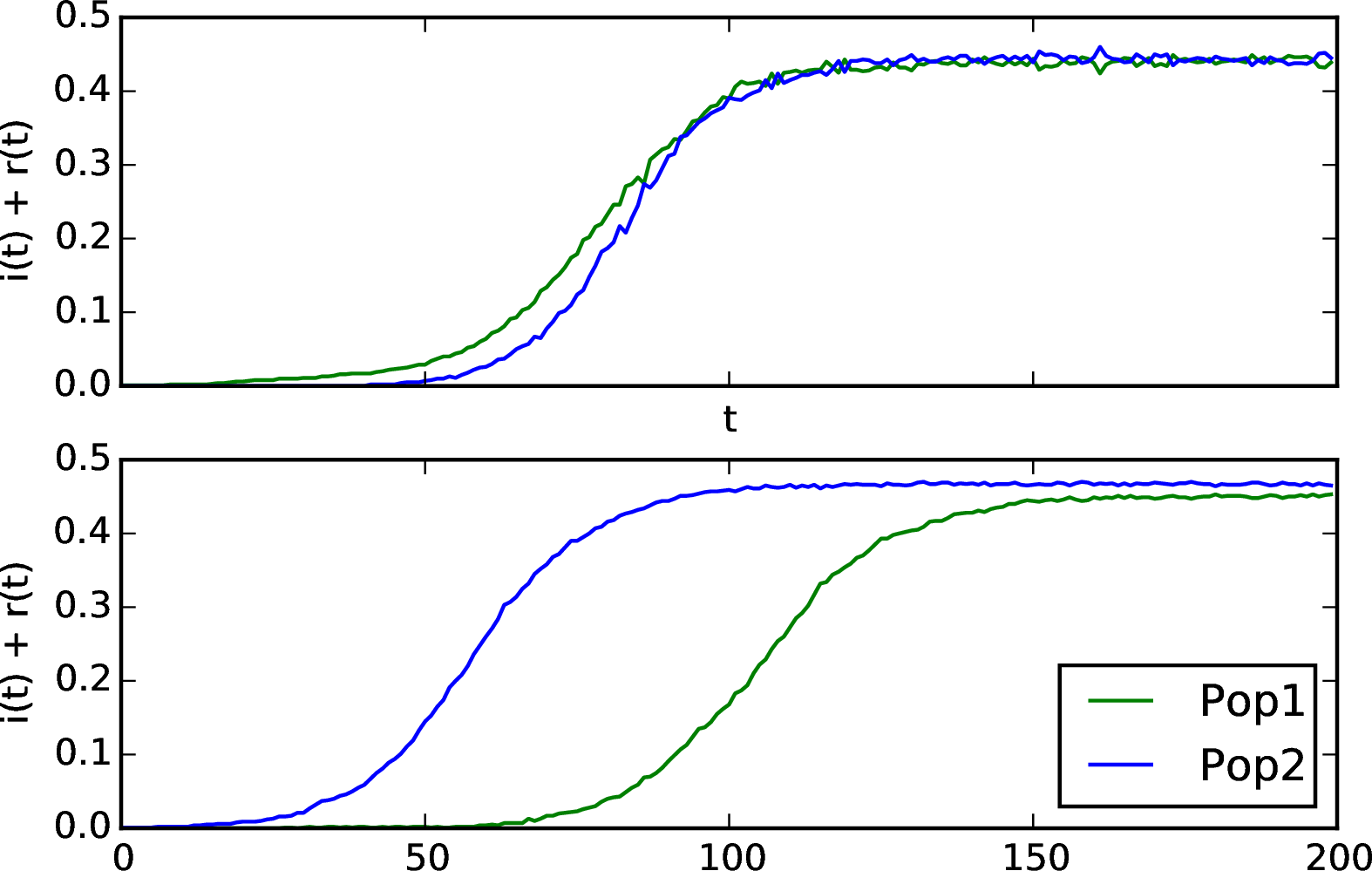}
\caption{\label{fig:simple}
A simple example with two metapopulations composed by people who stay always in their own metapopulation and a fraction $f$ of people who move between them randomly. In the top figure $f=0.1$, while in the bottom figure $f=0.01$. The outbreak dynamics in the second case are less synchronized.}
\end{center}
\end{figure}

As an illustrative case, we simulate a synthetic model. In Fig.~\ref{fig:simple} we show the total number of infections since the beginning of the simulation for two metapopulations, in two specific contexts. Individuals are equally assigned to either metapopulation and they belong to two classes: a fraction of people $(1-f)$ who do not travel out of their metapopulation, and a fraction of people $f$ who spend an equal amount of time, on average, in both. We use SEIR with the parameters mentioned before and we initialize the epidemics with a single infected case in one of the two metapopulations, chosen randomly. The top plot ($f=0.1$) shows a high level of synchronization, while the bottom plot ($f=0.01$) displays a clear delay in the growth of the epidemic size.

\begin{figure}[h!]
\begin{center}
\includegraphics[width=0.7\columnwidth]{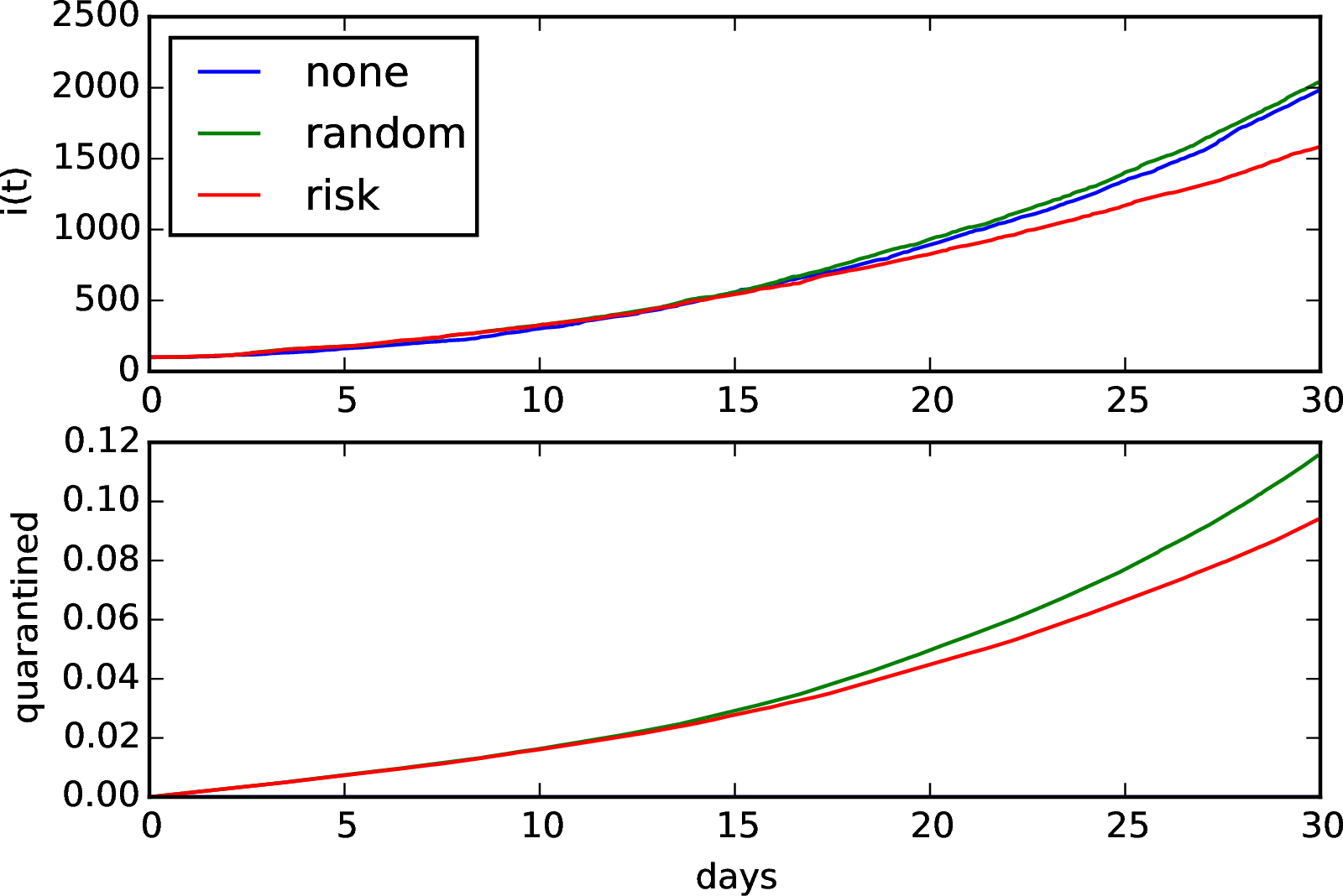}
\caption{\label{fig:results} The top plot shows how the total number of infected people changes in time when no countermeasures are taken (none), when people are quarantined randomly (random) and according to the highest risk rank (risk). The bottom plot shows how the number of people who have been put into quarantine grows in time. The proposed identification method reduces the number of infected individuals with fewer people in quarantine, using only aggregated information of the number of infected and mobility patterns from mobile phone data providers.   }
\end{center}
\end{figure}

We then test our approach initializing simulations with real-data, so that a single randomly chosen region is the unique source of infection with 100 cases. We use the first six months, from January to June 2013, to learn the movement habits of individuals. Then we perform simulations under the three scenarios mentioned above: no countermeasures, people quarantined randomly and people quarantined according to their risk rank. We set an adaptive quarantine rate of $\xi = \beta i(t)$ to match the countermeasure efforts with the speed of growth of the outbreak. Fig.~\ref{fig:results} shows results for the month of July 2013, in terms of how the global prevalence of the disease changes in time in the three cases. Despite the number of randomly quarantined people is pretty high at the end of the month (10\% of the population), it does not delay the spreading. Targeted quarantine based on risk, instead, manages to delay the spreading; at the end of the month there are 24\% fewer infected individuals than in the baseline cases.

\begin{figure}[h!]
\begin{center}
\includegraphics[width=0.7\columnwidth]{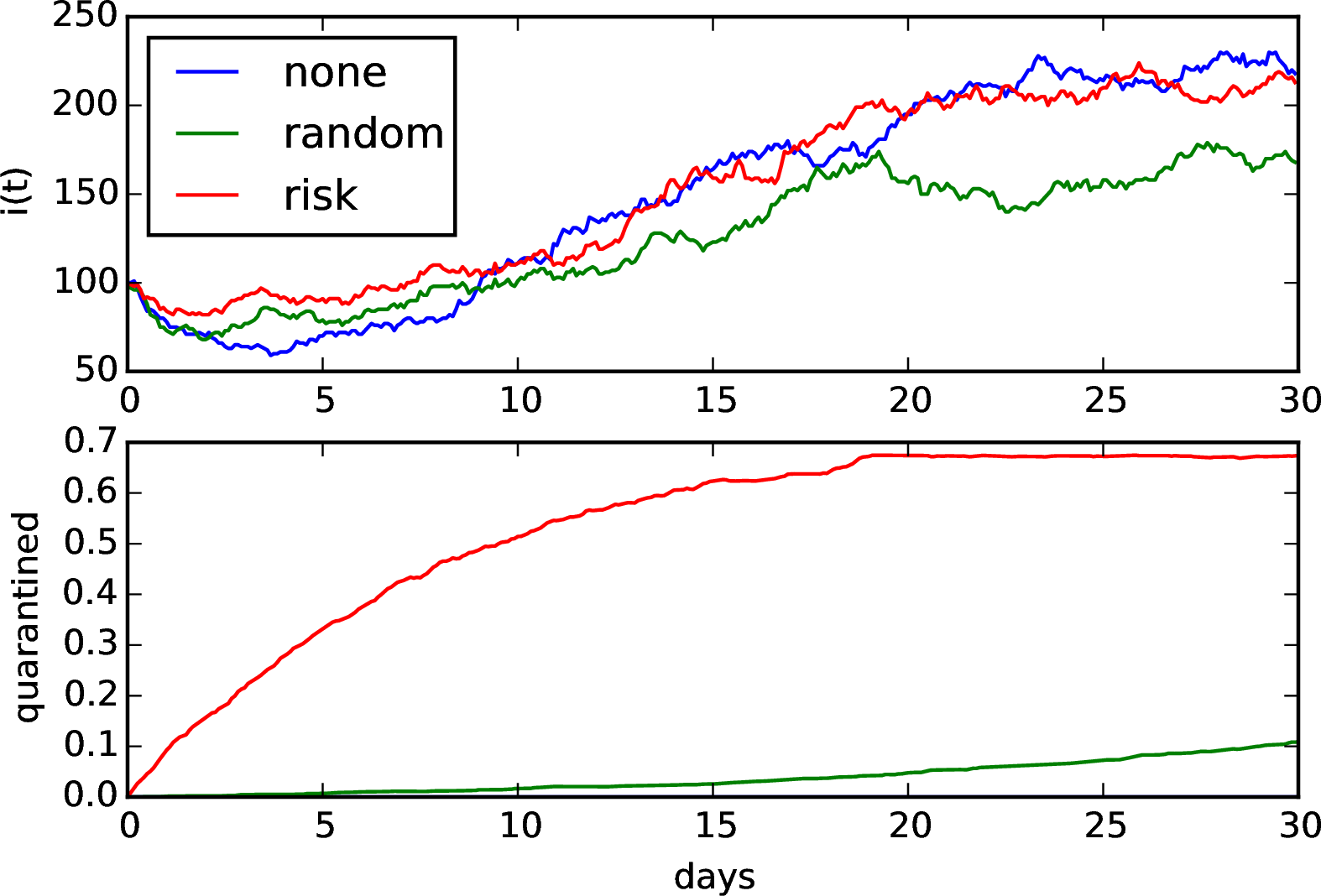}
\caption{\label{fig:infected-start} Number of infected (top plot) and quarantined (bottom plot) in the region where the first cases were initalized (hence, a region with higher risk than the others). Our proposed approach determines an increased number of infections in this region, while reducing the total aggregated number of infections.}
\end{center}
\end{figure}

\begin{figure}[h!]
\begin{center}
\includegraphics[width=0.7\columnwidth]{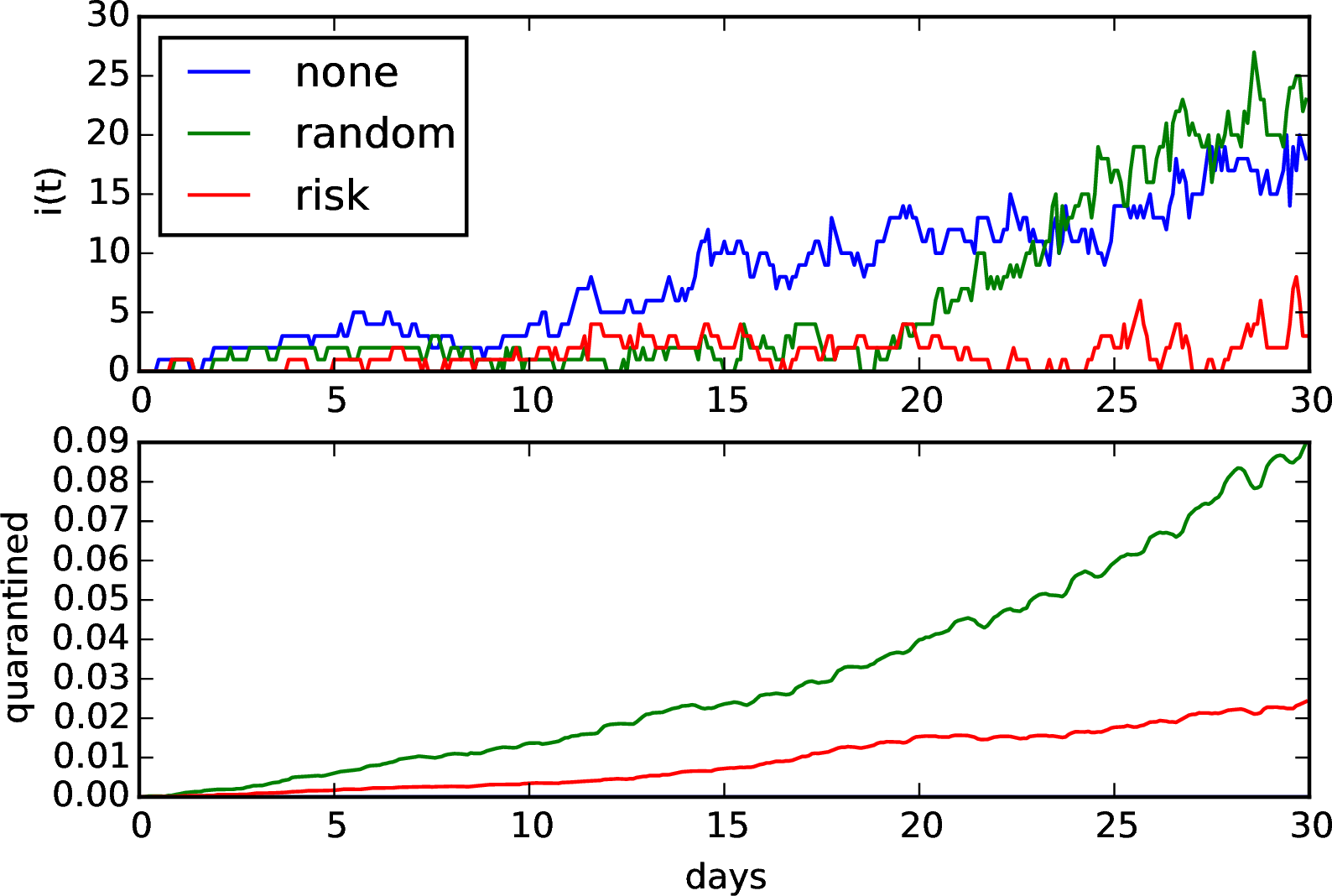}
\caption{\label{fig:infected-other} Number of infected (top plot) and quarantined (bottom plot) in a low-risk region. Our proposed approach determines a decreased number of infections in regions that have been less affected by the epidemic, such as the one shown here; this determines a delay in how the global number of infected grows in time.}
\end{center}
\end{figure}

This effect is obtained by restricting individuals who are in the areas with higher risk, specifically those who travel to low risk areas. This determines an increased number of infection cases in high-risk areas, as shown in Fig.~\ref{fig:infected-start} and a decreased number of infection cases in low risk areas, as shown in Fig.~\ref{fig:infected-other}.

\section{Discussion and Limitations}

This model assesses risk using data collected from mobile phones, hence it excludes people who do not use the mobile phones or share them with others. Since mobile penetration rates are already high and increasing in the vast majority of countries, including developing countries, we believe that the coverage problem will fade out as time goes. Another potential problem when dealing with network-data is the sparsity of the call activity, but recent studies try to overcome this limitation~\cite{Leontiadis_2014} by interpolating information in space and time. Furthermore, we would like to remark that the goal of this method is not to find \textit{every} high-risk individuals, but a \textit{large proportion} of them, given the data available. Moreover, it is worth noting that this method might be also be combined with other existing disease prevention and containment techniques already in use, such as contact tracing.


The model described in this paper requires access to sensitive data about individual call and mobility patterns. It is very important to take into account ethical and legislative issues arising from the use of these highly personal data. However, solutions based on the analysis of mobile data, such as that presented in this work, can play a critical role during emergencies.
For this reason, we believe it is acceptable to use such system when the benefits exceed the risks. We envision the use of such a system only in well-defined circumstances, within specific time intervals and geographic boundaries, within the limits defined by the law and under user informed consent.  The model could also be used to design a system that informs users only the users themselves about their own behavior, evaluating their the risk level and, potentially, suggesting them appropriate actions tailored to their risk profile (e.g., get tested, seek help, change lifestyle habits, etc.).

It is important to emphasize that a real-world system that would use this model would require access to two types of data: global information about the outbreak, which is already available  (e.g., estimated number of infected people in various geographic regions); and individual information about user mobility, which is of sensitive nature. For this reason, centralized deployments of the system might not be realizable, as user mobility might be unaccessible under local regulations and laws. On the other hand, decentralized deployments are to be preferred. In such deployments the mobility data are only stored on the devices. The risk-profile is calculated on the phone and it is shown only to the user, who might optionally choose to follow tailored advices.

Moreover, we want to stress that the model we propose simply gives a risk measure to each user. While this measure can be used to select who should be quarantined (i.e., the scenario we used for our simulation), it can also be used as a basis for less invasive measures, such as deciding who gets vaccination or who needs to see a doctor.

Finally, it is worth noting that technical and practical constraints might reduce the efficacy of mobility-based risk-assessment. In particular, we evaluate the model on mobility traces that correspond to an epidemics-free case. People might change significantly their mobility behavior once they are aware of the epidemic~\cite{Meloni2011}. Users might not carry their device with them at all times, hence making mobility traces and risk-assessment less effective. Mobility containment and other individual-based strategies are difficult to enforce and they heavily depend on the ultimate choices of individuals of accepting the recommendations.

\section{Related Work}

Human behavior can have a significant impact on infective disease dynamics. In turn, a complex interplay of disease spread, awareness of the disease, and population beliefs affect human behavior~\cite{Funk2010}. The mobility of a person, whether that person is infected or not, is a particularly important factor of disease spread~\cite{Rizzo2014}. Awareness-induced changes in movement patterns, such as a decision to avoid unsafe infected areas, often have a detrimental effect and might lead to even higher disease spreading, since they result in bringing the infection into previously isolated communities~\cite{Wang2012,Meloni2011}. At the same time, international travel restrictions have been shown to have a limited impact on disease spreading, due to the high heterogeneity of human mobility patterns~\cite{bajardi2011human}. In fact, it is this heterogeneity, both in terms of population behavior and a-priori infections, that drives disease development. In her discussion of HIV and other STDs transmission Aral argues that bridge groups, such as truckers, the police and the military personnel, transmit infections from highly infected groups, e.g., sex workers, to previously uninfected populations~\cite{Aral2000}. Our work is founded on the above observation, and we propose a model that explicitly takes the transmission of risk into account. While previous models consider artificial simulations~\cite{Buscarino2014} and long-distance~\cite{Merler2010} or multiscale~\cite{balcan2009multiscale} mobility networks in order to quantify possible outcomes of different metapopulations movement patterns on disease spread, we build our model upon individual mobility and interactions, as recorded by fine-grain cellular network traces.

Our work relies on mobile phone call records for estimating risk transfer in a population. The suitability of CDRs for tracking population movements and identification of spatial events in populations has been shown by Bengtsson et al.~\cite{Bengtsson2011} and Candia et al.~\cite{Candia2008}. Furthermore, when it comes to infectivity modeling, in~\cite{Eames2009} Eames et al. show that simple interaction potential measures, such as the total number of a user's connections (total degree), perform almost as well as more complex measures of interaction, such as individually weighted links. In further work the total node degree might be used to approximate a user's potential for contact.
Finally, in this work we do not modify the interaction network over time. Such modifications, orthogonal to our approach, are discussed in~\cite{Kamp2010}, and can be accounted for by having a time-dependent contact network.

\section{Conclusions}
In this paper we have proposed Progmosis, an approach to disease prevention and containment that goes beyond traditional epidemic modeling and contact tracing, and leverages behavioral data generated by mobile carrier networks to evaluate contagion risk on a per-user basis. The individual risk represents the loss incurred by not isolating or treating a specific person, both in terms of how likely it is for this person to spread the disease as well as how many secondary infections it will cause. We have developed and released an open-source tool that calculates this risk based on cellular network events. We have also simulated a realistic epidemic scenario, based on an Ebola virus outbreak. We have found that gradually restricting the mobility of a subset of individuals, selected using Progmosis greatly reduces the number of infected people, compared to a random choice.

This work focuses on a theoretical model and not on its actual translation into a real-world system. In particular, a decentralized deployment that preserves user privacy is to be preferred to a centralized one, which would require access to sensitive user data.  While computer-based simulations show promising results, they are obtained under specific assumptions; real-world constraints and challenges might greatly affect the effectiveness of this model. It is worth remarking that simulations were performed using data of a country that is currently Ebola-free according to WHO. Finally, we would also stress the fact that this work has not been commissioned neither by Orange nor by any other organisation for preparation to a real-world disease outbreak.

\bibliography{full_article-flat.bib}

\end{document}